\documentclass[aps,prc,preprint,superscriptaddress,showpacs]{revtex4}
\usepackage{graphicx}
\begin{document}

\preprint{Preliminary version}

\title{Neutron halo and the impact-parameter dependence of the 
$\pi^+ - \pi^-$ asymmetry in heavy ion collisions}

\author{Piotr Paw{\l}owski}
\email[Electronic address: ]{piotr.pawlowski@ifj.edu.pl}
\affiliation{Institute of Nuclear Physics, PL-31-342 Cracow, Poland }

\author{Antoni Szczurek}
\email[Electronic address: ]{antoni.szczurek@ifj.edu.pl}
\affiliation{Institute of Nuclear Physics, PL-31-342 Cracow, Poland }
\affiliation{University of Rzesz\'ow, PL-35-959 Rzesz\'ow, Poland}

\date{\today}

\begin{abstract}
The theoretical nuclear physics predictions and experimental results
on antiprotonic atoms lead
to different proton and neutron spatial distributions.
At CERN SPS energies production of positive and negative
pions differs for $pp$, $pn$, $np$ and $nn$ scattering.
These two facts lead to an impact parameter dependence
of the $\pi^+$ to $\pi^-$ ratio in $^{208}Pb + ^{208}Pb$ collisions.
A recent experiment at CERN seems to confirm qualitatively these predictions.
The results for two models considered are almost identical.
It may open a new possibility for determination of neutron density
distribution in nuclei.
\end{abstract}

\pacs{13.85.Ni,21.10.Gv,27.75.Dw}

\maketitle

The whole traditional microscopic nuclear physics is based
on the assumption that nuclei are built of protons and neutrons.
The so-called y-scaling analysis seems to confirm
that this is good assumption \cite{y_scaling}.

Electron scattering off nuclei provides direct information about
charge distribution (see for instance \cite{Frois77}), which is
closely related to the spatial distribution of protons.
The information about the neutron spatial distribution is not
accessible directly.

Experimentally one usually determines the difference $\Delta r_{np}$
between the neutron and proton mean square radii. However, this does not
allow one to make a difference between the neutron skin or neutron halo
model \cite{Trzcinska}. The radiochemical method applied to
antiprotonic atoms \cite{radiochemical1, radiochemical2, radiochemical3} allows one to measure
the neutron-to-proton ratio at the peripheries of nuclei. This measurement
suggests a density profile of the halo type, where half-radii
for neutron and proton densities are equal, and distributions
differ only in their diffuseness.

Another method of investigating neutron distribution is based on
the analysis of X-ray spectra of antiprotonic atoms \cite{Trzcinska}.
It was also suggested in \cite{DDS89} that parity violating electron
scattering could measure neutron density (for a newer review see
\cite{HPSM01}).

On the theoretical side the difference between the proton and
neutron distributions can be obtained in the framework
of Hartree-Fock (HF) method (see for example \cite{HL98})
or Hartree-Fock-Bogoliubov method (see for example
\cite{HFB_nuclear_skins}). For heavy nuclei one usually uses
zero-range nucleon-nucleon interaction of the Skyrme type.
Reliable neutron densities can be also obtained from a global analysis
of intermediate-energy elastic proton-nucleus scattering \cite{SH94,CKH03}.

Stable heavy nuclei exhibit an excess of neutrons over protons.
A recent experiment at CERN \cite{Rybicki}
for charged pion production in the $^{208}Pb + ^{208}Pb$ collision
has observed an interesting dependence of the ratio
$\frac{\text{d}\sigma}{\text{d}x_F}^{\!\!\pi^+}$/%
$\frac{\text{d}\sigma}{\text{d}x_F}^{\!\!\pi^-}$
on the Feynman variable $x_F$.
Surprisingly, the ratio for central and peripheral collisions
differs significantly \cite{Rybicki}.

\begin{figure}[htb] 
\begin{center}
\includegraphics[width=8cm]{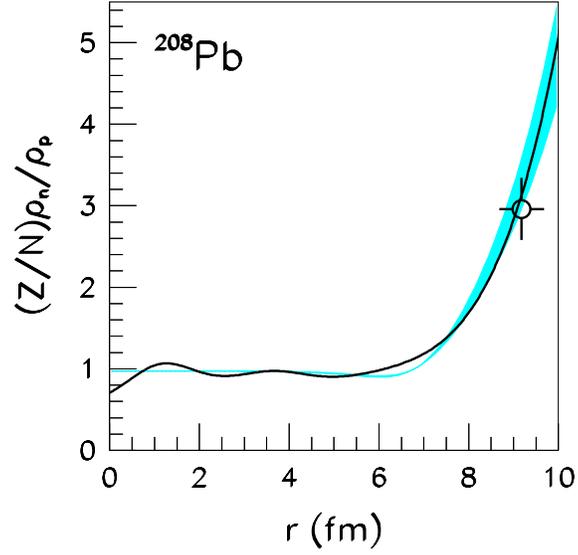}
\caption{
Neutron-to-proton ratio  $ \frac{Z}{N}
\frac{\rho_n(r)}{\rho_p(r)}$
for $^{208}Pb$.
The experimental point is taken from \cite{Trzcinska} while the experimental
grey band from \cite{Schmidt}. The result of the HFB calculation
\cite{HFB_nuclear_skins} is shown by the thin solid line.
\label{fig_rhon_rhop}
}
\end{center}
\end{figure}

In Fig.\ref{fig_rhon_rhop} we present the ratio $\rho_n(r) / \rho_p(r)$
as obtained from the HFB densities of $^{208}Pb$ \cite{HFB_nuclear_skins}.
The experimental point, taken from Ref.\cite{Trzcinska}, is a result of
interpolation of results obtained for other nuclei (the radiochemical
method can be used only for a limited set of elements).
The grey area shows a direct estimate of $\rho_n(r) / \rho_p(r)$
obtained from X-ray spectra of antiprotonic Pb atoms. This is slightly
model dependent and was obtained under assumption of Fermi-like
density profiles of halo type.
At large radii the HFB calculations are consistent with
these experimental results.

In the following we assume that pions are produced in elementary
nucleon-nucleon collisions.
Without loosing more generality we can write
\begin{equation}
\frac{d \sigma^{A_1 A_2 \rightarrow \pi^{\pm}}}{d x_F}(b,x_F;W) =
\sum_{\alpha \beta = p,n} N_{\alpha \beta \rightarrow \pi^{\pm}}(b;W) \;
\frac{d \sigma^{\alpha \beta \rightarrow \pi^{\pm}}}{d x_F}(x_F;W)
\; ,
\label{NN_scattering}
\end{equation}
where $W$ is energy per binary nucleon-nucleon collision and
$N_{\alpha \beta \rightarrow \pi^{\pm}}$ are numbers
of collisions of a given type.
Only elementary cross sections for the $pp \rightarrow \pi^{\pm}$
processes are known experimentally.
Therefore in the following we shall use elementary cross sections
$\frac{d \sigma}{d x_F} (x_F;W)$ calculated in the HIJING model \cite{HIJING}.
We shall leave the problem of the quality of the model 
for a separate detailed analysis.

\begin{figure}[htb] 
\begin{center}
\includegraphics[width=14cm]{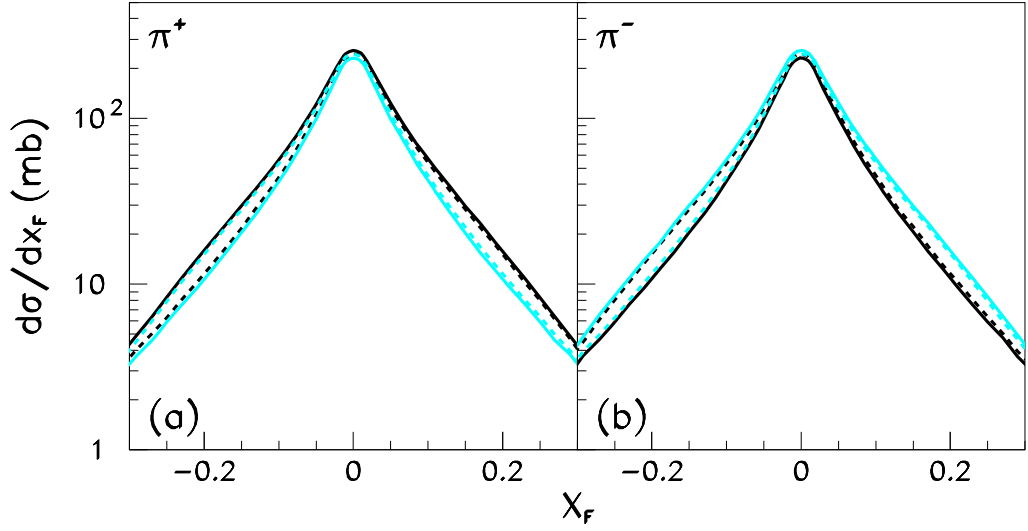}

\caption{
$\frac{d \sigma}{d x_F}$ as a function of Feynman $x_F$
for $\alpha \beta \rightarrow \pi^+$ (left panel)
and $\alpha \beta \rightarrow \pi^-$ (right panel)
for different subcollisions: $pp$ (black solid), $nn$ (grey solid),
$pn$ (black dashed) and $np$ (grey dashed).
\label{fig_pip_pim}
}
\end{center}
\end{figure}

In Fig.\ref{fig_pip_pim} we show $x_F$ dependence of
the cross section for different types of collisions. We have used
$\sigma_{ine}^{NN}$ = 30 mb to normalize all elementary cross
sections.
A clear difference between different cross sections can be seen.
This asymmetry in elementary collisions in conjunction with
the asymmetry of proton and neutron spatial distributions should
lead to asymmetries in nuclear collisions which will be discussed
in the present note.
Only four elementary cross sections are independent.
The cross sections for the $\pi^-$ production (right panel)
can be obtained from those for the $\pi^+$ production (left panel)
thanks to isospin symmetry. In addition at $x_F >$ 0 one has
$\sigma_{pp \to \pi^{\pm}} \approx \sigma_{pn \to \pi^{\pm}}$
and
$\sigma_{nn \to \pi^{\pm}} \approx \sigma_{np \to \pi^{\pm}}$.

When calculating the ratio of $\pi^+$ to $\pi^-$ cross sections
it is sufficient to know the fractions
$f_{\alpha \beta} = \frac{N_{\alpha \beta \to \pi^{\pm}}}
{\sum_{\alpha \beta} N_{\alpha \beta \to \pi^{\pm}}}$.
It is far from obvious how to calculate the fractions $f_{\alpha \beta}$.
In the most naive approach
\begin{equation}
f_{\alpha \beta} = \frac{N_{\alpha/1} N_{\beta/2}}
{\sum_{\alpha \beta = p, n} N_{\alpha/1} N_{\beta/2}} \; ,
\label{fraction_0}
\end{equation}
where $N_{\alpha/1}$, $N_{\beta/2}$ are fractions of protons/neutrons
in the nucleus $A_1$ and $A_2$, respectively.
In this simple model there is neither impact parameter nor energy
dependence of $f_{\alpha \beta}$, i.e. as a consequence of
Eq.(\ref{NN_scattering}) the ratio
\begin{equation}
R_{+/-}(x_F,b;W)
\equiv \frac{d\sigma^{A_1 A_2 \rightarrow \pi^+}(x_F,b;W)/dx_F}
            {d\sigma^{A_1 A_2 \rightarrow \pi^-}(x_F,b;W)/dx_F}
\label{plus_minus_ratio}
\end{equation}
depends only on the Feynman $x_F$ and center-of-mass energy.
Recent preliminary experimental results of the NA49 collaboration
\cite{Rybicki} seem to contradict this simple approach.

It is often assumed that the dynamics of nuclear collisions
is governed by the number of binary collisions \cite{binary_collisions}.
The number of binary nucleon-nucleon collisions at a given impact
parameter is proportional to the nucleus-nucleus thickness
\footnote{In the present analysis we take zero-range NN interaction
to avoid introduction of new not well known parameters.
The effect of finite-range interaction will be discussed elsewhere.}
\begin{equation}
T_{A_1 A_2}(\vec{b}) =
\int d^2 s_1 T_{A_1}(\vec{s}_1) \; T_{A_2}(\vec{s}_1 - \vec{b}) =
\int d^2 s_2 T_{A_1}(\vec{s}_2 - \vec{b}) \; T_{A_2}(\vec{s}_2) \; .
\label{coll_thick}
\end{equation}
In Eq.(\ref{coll_thick}) we introduced
$T_{A_i}(\vec{b}) = \int d z_i \; \rho_{A_i}(\vec{b},z_i)$,
where $\rho_{A_i}$ is the density function of the nucleus $A_i$
normalized to the number of nucleons.
Analogously the number of binary collisions of a given type
in Eq.(\ref{NN_scattering}) can be written as
\begin{equation}
N_{\alpha \beta \rightarrow \pi^{\pm}}(b; W) =
T_{A_1 A_2}^{\alpha \beta}(\vec{b}) \cdot \sigma_{ine}^{NN}(W) \; ,
\label{fraction_binary}
\end{equation}
where
\begin{equation}
T_{A_1 A_2}^{\alpha \beta}(\vec{b}) =
\int d^2 s_1 T_{A_1}^{\alpha}(\vec{s}_1)
          \; T_{A_2}^{\beta}(\vec{b}-\vec{s}_1) =
\int d^2 s_2 T_{A_1}^{\alpha}(\vec{b}-\vec{s}_2)
          \; T_{A_2}^{\beta}(\vec{s}_2)
\; ,
\label{generalized_coll_thick}
\end{equation}
where now $T_{A_i}^p$ and $T_{A_i}^n$ are nucleus thicknesses of
protons and neutrons, respectively.
We have assumed one universal inelastic cross section
$\sigma_{ine}^{\alpha \beta}(W) = \sigma_{ine}^{NN}(W)$.
When properly normalized,
Eq.(\ref{generalized_coll_thick}) describes the probability
of the binary $\alpha \beta$ collisions and is a simple
generalization of Eq.(\ref{coll_thick}). Of course
\begin{equation}
T_{A_1 A_2}(\vec{b})
 = \sum_{\alpha \beta} T_{A_1 A_2}^{\alpha \beta}(\vec{b}) \; .
\label{binary_sum_rule}
\end{equation}

As a second limiting case we shall consider the wounded nucleon model
\cite{wounded_nucleon}. We shall assume that the production
of particles is proportional to the number of wounded nucleons
rather than to the number of binary collisions.
The simplest version of the model must be generalized to
include different distributions of protons and neutrons
and differences of elementary cross sections.
Somewhat schematically the cross section for the nuclear collision
can be written as
\begin{eqnarray}
 \frac{d\sigma ^{A_{1}A_{2}\rightarrow \pi ^{\pm}}}{dx_{F}}(b,x_{F};W)
\propto \sum _{\alpha ,\beta =p,n}
\left[N_{\alpha /A_{1}}^{wou}(b,W)\; w_{2}^{\beta }(b)\;
 \frac{d\sigma ^{\alpha \beta \rightarrow \pi ^{\pm }}}{dx_{F}}(x_{F};W)\right.
 \nonumber
   \\
+\left. N_{\alpha /A_{2}}^{wou}(b,W)\; w_{1}^{\beta }(b)\;
  \frac{d\sigma ^{\beta \alpha \rightarrow \pi ^{\pm
      }}}{dx_{F}}(x_{F};W)\right]  \; .  
 \label{wounded_nucleon_model}
\end{eqnarray}
In the formula above $N_{\alpha/A_i}^{wou}$ is the number of
wounded $\alpha$ (p or n) in nucleus $A_1$ or $A_2$
and $w_i^{\beta}$ is the probability that the wounded $\alpha$
interacted with $\beta$ (p or n) from nucleus $A_2$ or $A_1$,
respectively.
Eq.(\ref{wounded_nucleon_model}) is equivalent to
Eq.(\ref{NN_scattering})
with
\begin{equation}
N_{\alpha \beta \rightarrow \pi^{\pm}} =
      N_{\alpha/A_1}^{wou}(b,W) \; w_2^{\beta}(b)
    + N_{\beta/A_2}^{wou}(b,W) \; w_1^{\alpha}(b) \; .
\label{fractions_wounded}
\end{equation}

By construction, in our approach the numbers of wounded protons
and neutrons reproduce the well known formula from \cite{wounded_nucleon}
for the number of wounded nucleons
\begin{equation}
N_{N/A_i}^{wou}(b) = N_{p/A_i}^{wou}(b) + N_{n/A_i}^{wou}(b) \; .
\label{N_wounded_sum}
\end{equation}
Numbers of wounded protons and neutrons were calculated
in the zero-range nucleon-nucleon approximation.
In calculating the number of wounded nucleons we have taken
$\sigma_{ine}^{NN}$ = 30 mb, relevant for CERN SPS energies.
Our construction requires also
\begin{equation}
w_i^p(b) + w_i^n(b) = 1 \; .
\label{N_wounded_fraction_sum}
\end{equation}
In this communication the fractions $w_i^{\alpha}$ were estimated as
\begin{equation}
w_i^{\alpha}(b) = \frac{N_{\alpha/A_i}^{wou}(b)}
{N_{p/A_i}^{wou}(b) + N_{n/A_i}^{wou}(b)} \; ,
\label{w_i}
\end{equation}
which by construction fulfils (\ref{N_wounded_fraction_sum}). With
this prescription Eq.(\ref{fractions_wounded}) reduces to:
\begin{equation}
N_{\alpha \beta \rightarrow \pi^{\pm}}
 = \left( N_{N/A_1}^{wou}(b,W)+N_{N/A_2}^{wou}(b,W) \right)
 \; w_1^{\alpha}(b) \; w_2^{\beta}(b) \; .
\label{fractions_weights_wounded}
\end{equation}
%

In the present note we use proton and neutron densities
calculated in the Hartee-Fock-Bogoliubov method 
\cite{HFB_nuclear_skins}
with Skyrme interaction SLy4 (see Fig.\ref{fig_rhon_rhop}).

Above we have only sketched the model. More details
will be presented and discussed in detail elsewhere \cite{PS_paper}.

\begin{figure}[htb] 
\begin{center}
\includegraphics[width=14cm]{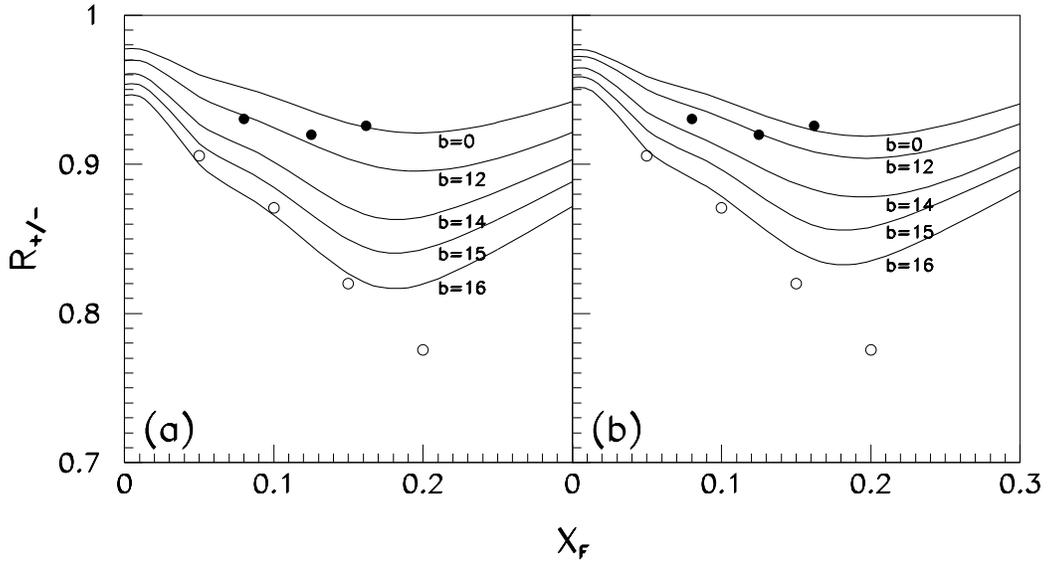}
\caption{
The ratio $R_{+/-}$ as a function of the Feynman $x_F$
for the model with binary collision (left panel)
and for the model with the scaling with number of wounded
nucleons (right panel). Also shown are the preliminary
experimental data from Ref.\cite{Rybicki}.
\label{fig_R_xf}
}
\end{center}

\end{figure}

In Fig.\ref{fig_R_xf} we present the ratio $R_{+/-}$
(see Eq.(\ref{plus_minus_ratio}))
as a function of $x_F$ for different values of the impact parameter $b$
in the two models considered.
For comparison we show
preliminary experimental data for "central collisions" (solid circles)
and "peripheral collisions" (open circles) presented in \cite{Rybicki}.
The notion of central and peripheral collisions was not specified
in \cite{Rybicki}. Therefore the data can be used only as an indication
of the effect. At present no quantitative comparison of model results
to experimental data is possible.
Our approach explains the experimental data provided they are
extremely peripheral (this requires their more detailed analysis).
Our results are reliable up to $x_F \approx$
0.2. It is known that the HIJING code does not describe the
$\pi^+$-$\pi^-$ asymmetry in elementary collisions in the region
of large $x_F$.

\begin{figure}[htb] 
\begin{center}
\includegraphics[width=14cm]{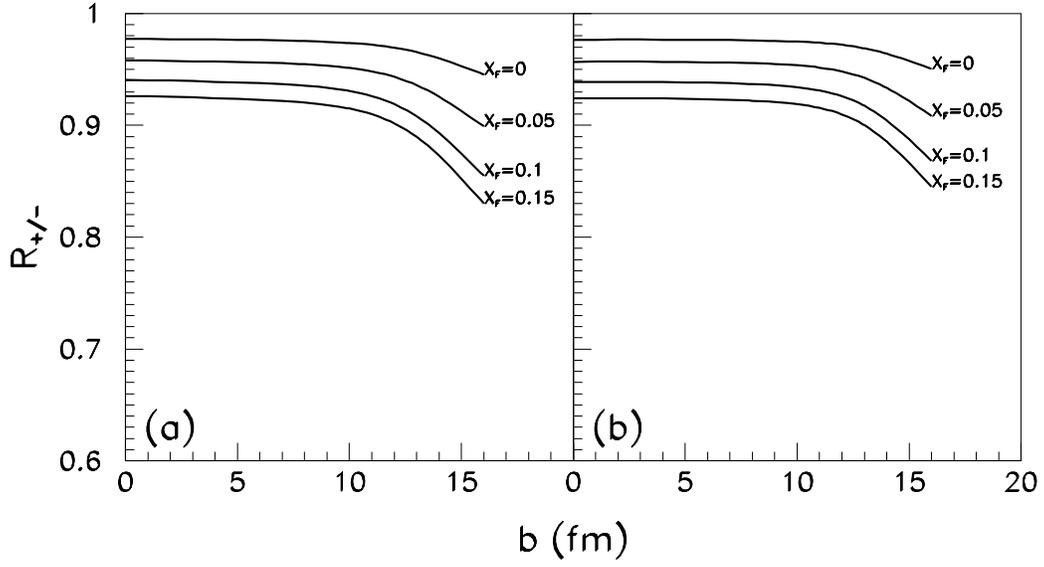}
\caption{
The ratio $R_{+/-}$ as a function of impact parameter
for the model with binary collision (left panel)
and for the model with the scaling with number of wounded
nucleons (right panel).
\label{fig_R_b}
}
\end{center}
\end{figure}

For completeness in Fig.\ref{fig_R_b} we present the ratio
$R_{+/-}$ as a function of the impact parameter $b$ for different
values of $x_F$.
Surprisingly the binary collision picture gives very similar results
to the predictions of the wounded nucleon model.

The predictions of the two models are almost identical.
This suggests that a detailed comparison of model results
with the well defined ($x_F$, $b$) experimental data
could open a new possibility to study the neutron density
profile. We expect that the NA49 collaboration
at CERN will be able to gather the corresponding experimental data
in the near future. Can it provide a method competitive
to that offered by proton-nucleus elastic scattering, antiprotonic atoms
or parity violating electron scattering? Of course results of
these methods must finally converge. Therefore one may hope that
together they will provide more reliable information on
neutron distribution in nuclei.

\begin{acknowledgments}
We are indebted to Andrzej Rybicki for inspiring this analysis,
Agnieszka Trzci\'nska for a discussion about
antiprotonic atoms and Jacek Dobaczewski for providing us
the nucleon distribution functions from their HFB calculations.
\end{acknowledgments}

\bibliography{n_skin_rev}

%



\end{document}